\documentclass[journal=jpclcd,
               layout=twocolumn,
               layout=onecolumn,
               manuscript=letter
               ]
               {achemso}

\usepackage[version=3]{mhchem} 
\usepackage{xcolor}
\usepackage{graphicx}

\author{Jing Sun}
\email{jing.sun@pci.uni-heidelberg.de}
\affiliation[Heidelberg University]
{Theoretische Chemie, Physikalisch-Chemisches Institut,
Universität Heidelberg,\\ 69120 Heidelberg, Germany}
\author{Oriol Vendrell}
\email{oriol.vendrell@uni-heidelberg.de}
\affiliation[Heidelberg University]
{Theoretische Chemie, Physikalisch-Chemisches Institut,
Universität Heidelberg,\\ 69120 Heidelberg, Germany}

\title[Chemical rates in cavities]
  {On the Suppression and Enhancement of Thermal Chemical Rates in a Cavity}

\begin{document}

\begin{tocentry}
    \includegraphics[width=5cm]{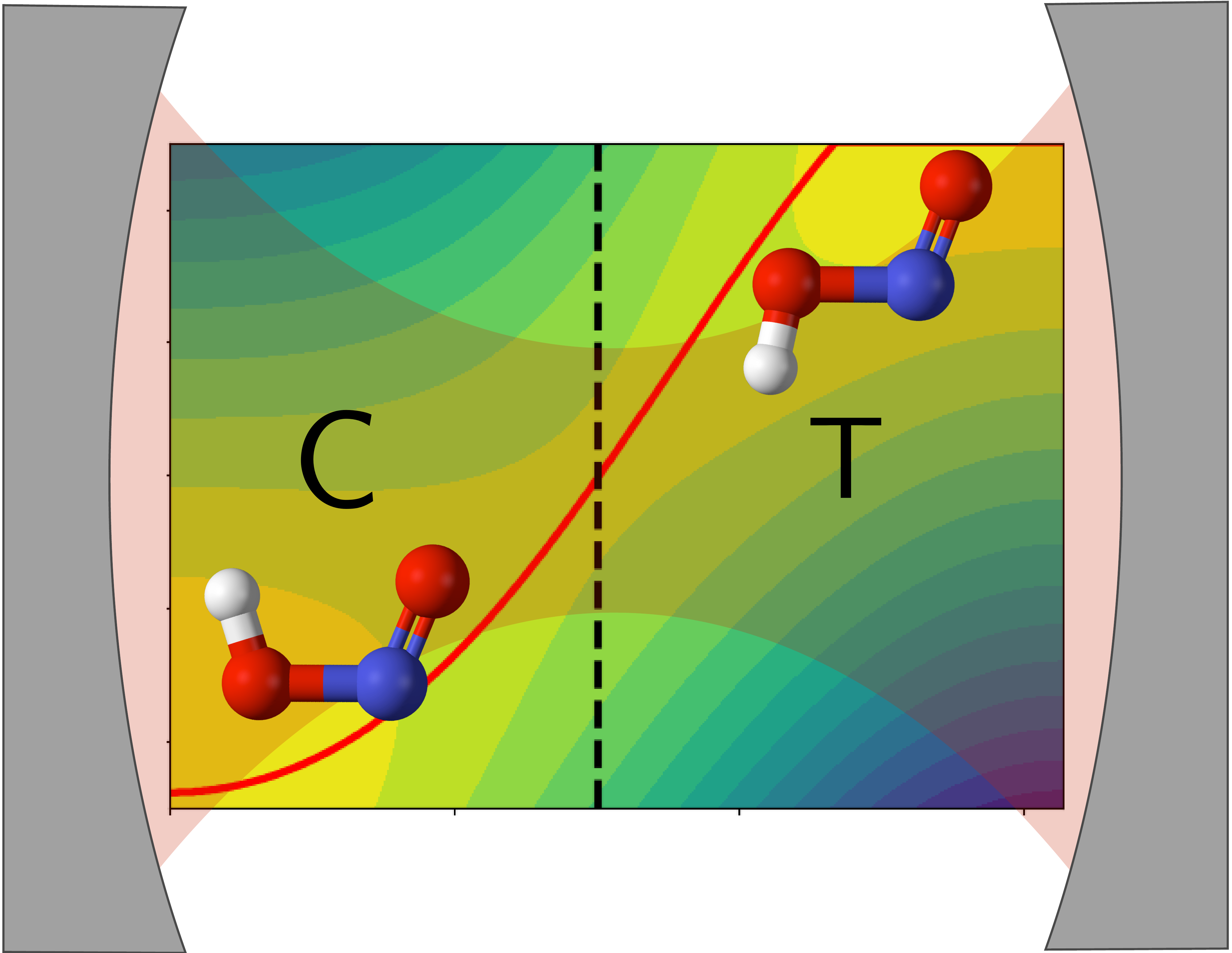}
\end{tocentry}




\begin{abstract}
  The observed modification of thermal chemical rates in Fabry-Perot cavities
  remains a poorly understood effect theoretically. Recent breakthroughs explain
  some of the observations through the Grote-Hynes theory, where the cavity
  introduces friction with the reaction coordinate, thus reducing the
  transmission coefficient and the rate. The regime of rate enhancement, the
  observed sharp resonances at varying cavity frequencies, and the survival
  of these effects in the collective regime remain mostly unexplained. In this
  paper, we consider the \emph{cis}-\emph{trans} isomerization of HONO
  atomistically using an \emph{ab-initio} potential energy surface. We evaluate the
  transmission coefficient using the reactive flux method and identify the
  conditions for rate acceleration. In the underdamped, low-friction regime of
  the reaction coordinate, the cavity coupling enhances the rate
  with increasing coupling strength until reaching the Kramers turnover point.
  Sharp resonances in this regime are related to cavity-enabled energy
  redistribution channels.
\end{abstract}



Vibrational strong coupling (VSC) has emerged as a very active front in the field of 
polaritonic chemistry since the pioneering demonstrations of Rabi splitting in the
infrared domain in Fabry-Perot configurations~\cite{hut_12_1624,sha_15_5981,ebb_16_2403}.
At the core of this sub-discipline lies the promise of modifying~\cite{dun_16_13504}, and ultimately
controlling~\cite{yan_20_919} the mechanisms and rates of thermal chemical reactions using the vacuum fields
of cavities~\cite{dun_18_965}.
Besides important breakthroughs in the linear and non-linear spectroscopy of VSC
systems~\cite{dun_16_13504,dun_18_965,yan_20_919,fas_21_11444},
the more spectacular results remain the experiments reporting the modification of chemical
rates in cavities by the Ebbesen group and
others~\cite{tho_16_11462,lat_19_10635,tho_19_615a,ver_19_15324,tho_20_249,imp_21_191103}.
These experiments have triggered the proposal of several theoretical models to explain
how the cavity modifies the ground electronic state structure~\cite{gal_19_021057}
and spectroscopy~\cite{pin_15_053040} and,
more recently, how it modifies reaction
rates~\cite{cam_19_4685,li_20_234107,vur_20_3557,li_21_1315,yan_21_9531,man_22_014101}.

Theoretical models based on the Grote-Hynes theory~\cite{hyn_86_149}
predict the suppression of the transmission coefficient with increasing cavity coupling
due to increased friction at the top of the reaction
barrier~\cite{li_21_1315,yan_21_9531,man_22_014101}.
How cavities can enhance chemical reactions~\cite{lat_19_10635}, how sharp
resonances of the cavity with vibrational modes affect the
mechanism~\cite{lat_19_10635,tho_19_615a}, and how these effects survive in the
collective VSC regime, have remained poorly understood questions. 

Here, we simulate the rate of a realistic isomerization reaction atomistically
using an \emph{ab initio} potential, both for one and several HONO molecules, in
the VSC regime.
Our simulations explain how the cavity enhances chemical rates
in the underdamped regime and capture the turnover from the underdamped to the damped regime as a
function of the cavity coupling strength.
Moreover, we explain how, in the underdamped regime, sharp
resonances of the cavity with vibrational modes can strongly affect the reaction
rate.
Finally, our results show how, in the collective VSC regime, the strong direct coupling to the
reaction coordinate ceases to be the determining factor in the cavity effect.

\begin{figure}[t]
\includegraphics[width=8cm]{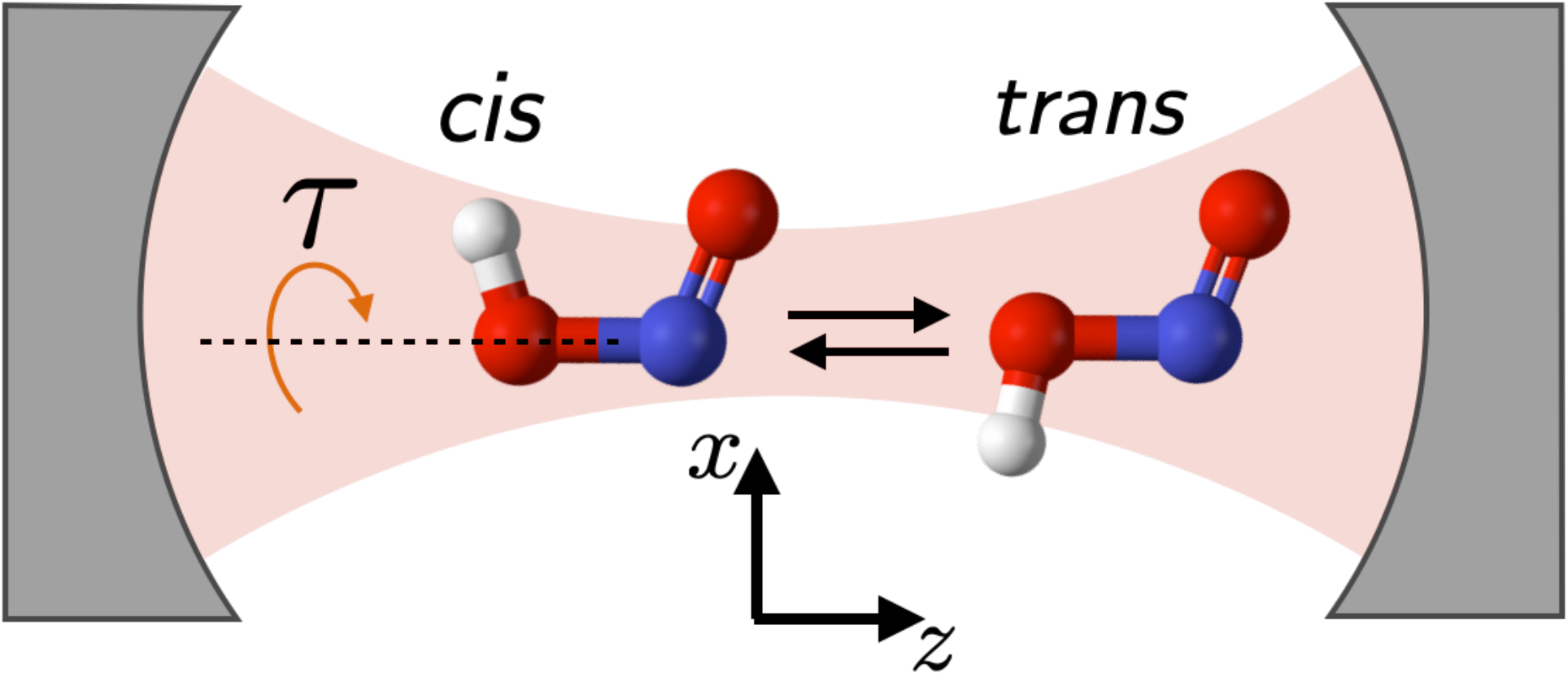}
\caption{
    \label{fig:hono}
    \emph{cis}-\emph{trans} isomerization reaction in HONO.
    The axes indicate the body-fixed frame of the molecules in the simulation.
    The presence of the cavity is indicated schematically and is not to scale.
    HONO is characterized by 6 vibrational coordinates:
    $3$ stretching modes, O$-$H, O$-$N and N$=$O; $2$ bending modes,
    H$-$O$-$N and O$-$N$=$O; $1$ torsion mode $\tau$,
the isomerization reaction coordinate.}
\end{figure}



Our starting point is the Hamiltonian for a molecular ensemble coupled to one or several cavity modes
\begin{align}
    \label{eq:Ham}
    \hat{H} & = \sum_{l=1}^{N}\hat{H}_{mol}^{(l)} + \hat{H}_{cav}
\end{align}
with
\begin{align}
    \label{eq:HamMol}
    \hat{H}_{mol}^{(l)} & = \sum_{j_l=1}^{F} \frac{\hat{P}_{j_l}^2} {2M_{j_l}}
     + \hat{V}(R_{1_l}\ldots R_{F_l}), \\
    \label{eq:HamCav}
    \hat{H}_{cav} &= \frac{1}{2}\left[ \hat{p}_{cav}^2 +\omega_{cav}^2
    \left( \hat{q}_{cav}+\frac{\boldsymbol{\lambda}}{\omega_{cav}}
    \cdot
    \sum^{N}_{l=1}\hat{\boldsymbol{\mu}}^{(l)})\right)^2 \right],
\end{align}
and where $V(R_{1_l}\ldots R_{F_l})$ denotes the ground electronic state
potential energy surface (PES) of the $l$-th molecule with momenta $P_{j_l}$ and
positions $R_{j_l}$.
Hence, the Born-Oppenheimer (BO) approximation is assumed within each molecule,
and
$\boldsymbol{\mu}^{(l)}\equiv \boldsymbol{\mu}^{(l)}(R_{1_l}\ldots R_{F_l})$
is the permanent dipole vector of the l-th molecule.
$\hat{H}_{cav}$ can be reached from the Coulomb-gauge light-matter interaction
Hamiltonian by taking the long wave approximation followed by a unitary
transformation to the length form~\cite{fli_17_3026,hau_21_094113,li_20_234107}.
For later convenience, we write $\hat{H}_{cav}$ in its position-momentum
representation ($\hat{q}_{cav}$, $\hat{p}_{cav}$). $\omega_{cav}$
corresponds to the cavity mode frequency.
This form of the light-matter interaction has become standard in most
theoretical studies of VSC~\cite{li_20_234107,sid_20_7525,li_21_1315,yan_21_9531},
and details on its derivation~\cite{pow_59_427,fli_17_3026}
and properties~\cite{sch_20_975} can be found elsewhere.
The parameter $\boldsymbol{\lambda}$ equals the coupling strength
$\lambda=\sqrt{1/\epsilon_0 V}$ times the unit polarization vector
$\boldsymbol{\epsilon}$ of the cavity mode, and $V$ represents the cavity
volume~\cite{fli_17_3026}.
Similarly to other studies and to facilitate comparisons, we introduce the
coupling parameter $g = \lambda \sqrt{\hbar\omega_{cav}/2}$, which has units of
electric field (using this relation and $\hat{q}_{cav} =
\sqrt{\hbar/2\omega_{cav}}(\hat{a}^\dagger+\hat{a})$, the linear coupling term
in $\hat{H}_{cav}$ reads $g\,
\boldsymbol{\epsilon}\cdot\boldsymbol{\mu}\,(\hat{a}^\dagger+\hat{a})$).

%
The simplicity of
the unimolecular reaction mechanism in HONO makes it an ideal benchmark system to
understand how dynamical cavity effects can modify chemical rates
as compared, e.g., to bimolecular
reactions in solution~\cite{tho_19_615a,cli_20_23545}.
We base our study on the CCSD(T)-quality \emph{ab initio} potential energy surface (PES)
of Richter et al.~\cite{ric_04_1306}, which features a
reaction barrier height of about 0.51~eV (49 kJ/mol) and where the \emph{trans}
isomer is 11~meV more stable than the \emph{cis} one.
Quantum dynamics studies of the HONO isomerization triggered by strong laser
pulses have been based on this PES~\cite{ric_04_6072,ric_07_164315}.
Despite its simplicity, this chemical reaction constitutes a fully coupled
and rich dynamical system. Similarly to other isomerization
reactions, e.g. involving hydrocarbons~\cite{mon_79_4056,ros_80_162,kuh_88_3261},
it takes place in the underdamped regime.
Throughout this work, the molecules are kept at a fixed orientation with
respect to the polarization direction of the cavity mode. In this way,
we focus on the coupling of the $\mu_x$ dipole component to the cavity polarizaton.
As shown in the SI, this component of the molecular dipole has the strongest modulation
at the isomerization barrier configuration.

The molecule-cavity system is considered within the classical approximation,
thus transforming all coordinate operators in Hamiltonians~\ref{eq:HamMol} and
\ref{eq:HamCav} to classical functions. A classical description of the VSC
regime is not new and has been successfully applied to bulk systems described by
force-field potentials~\cite{li_20_18324} and model Hamiltonians~\cite{li_21_1315,wan_21_}.
The \emph{cis}-\emph{trans} reaction rate
is described with the reactive flux method
for the classical rate
constant~\cite{mon_79_4056,ros_80_162,cha_87_,ber_88_3711,kuh_88_3261}
\begin{align}
    \label{eq:kt}
    K(t) &= x_{cis}^{-1}
    \langle \dot{\tau}(0)\,
            \delta[\tau(0)-\tau^{\ddag}]\,
            \theta[\tau(t)] \rangle,
\end{align}
where $x_{cis}$ is the equilibrium fraction of HONO at the \emph{cis}
geometry, $\dot{\tau}(0)$ is the initial velocity of a phase-space point
perpendicular to the dividing surface between reactants and products, and
$\tau^{\ddag}$ is the torsion angle corresponding to the transition state (TS)
geometry.
The brackets indicate the canonical ensemble average over trajectories,
where we considered a temperature of 300~K throughout. The Heaviside function $\theta[\tau]$
is defined to be one
for the \emph{trans} configurations, and zero otherwise.
The exact reactive flux is obtained in the limit $t\to\infty$, in practice when
the plateau for $K(t)$ is reached~\cite{ber_88_3711}.
This occurs when all classical trajectories
starting from the dividing surface become trapped at either the reactants or
products side.
For example, for the isomerization reaction of $n$-butane in the low-friction
environment of a van der Waals liquid this relaxation time is about
1~ps~\cite{ros_80_162}.
Now, since~\cite{cha_87_}
\begin{align}
    \label{eq:k0+}
    \lim_{t\to 0^{+}}K(t) = K_{TST},
\end{align}
one can introduce a transmission coefficient $\kappa(t) = K(t)/K_{TST}$
as the quotient of the numerically exact reactive
flux and the reactive flux without recrossing, i.e. the TST assumption.  
$K_{TST}$ can be evaluated conveniently using Eyring's equation~\cite{eyr_35_107,han_90_251},
while $\kappa(t)$ is obtained from classical trajectories.
As shown in the SI, and as has been discussed in other
works~\cite{li_20_234107,vur_20_3557},
$K_{TST}$ is, to a very good approximation, insensitive to cavity effects.
Therefore, we consider $K_{TST}$ to be completely
cavity-independent and describe the cavity effect on the rate as 
\begin{align}
    \label{eq:prodkappa}
    K_{cav} = \kappa_{cav} \kappa_0 K_{TST},
\end{align}
where $K_0 \equiv \kappa_0 K_{TST}$ is the formally exact classical rate
outside the cavity.
Here and in the following, transmission coefficients and rate constants
without a time argument refer to their plateau value.
Clearly, both
$\kappa$ and $\kappa_0$ lie in the $[0,1]$ range but $\kappa_{cav}$ can be both
larger or smaller than one, corresponding to a chemical rate enhancement or suppression,
respectively.

In the following, we theoretically demonstrate that both enhancement and
suppression of reaction rates are possible within a cavity for realistic chemical
processes. Although we rely on classical rate theory~\cite{han_90_251}, we note
that tunneling corrections for hydrogen abstraction reactions at 300~K result
in variations of the rate within the same order of
magnitude~\cite{mas_02_11760}. For reactions involving heavier elements, quantum
corrections to the rates are even more insignificant.
Along these lines, there is no reason to assume, a priori, that photonic modes
with frequencies similar to the atomic vibrations, and in thermal equilibrium,
shall result in significant quantum effects that affect the general
conclusions derived from classical rate theories for VSC systems.
This does not exclude situations where quantum effects may be important for
quantitative descriptions of cavity-modified rates in reactions involving light
atoms, as it is sometimes the case for rates outside
cavities~\cite{mil_79_6810,mat_98_4828,mas_02_11760}.
%

\begin{figure}[t]
\includegraphics[width=8cm]{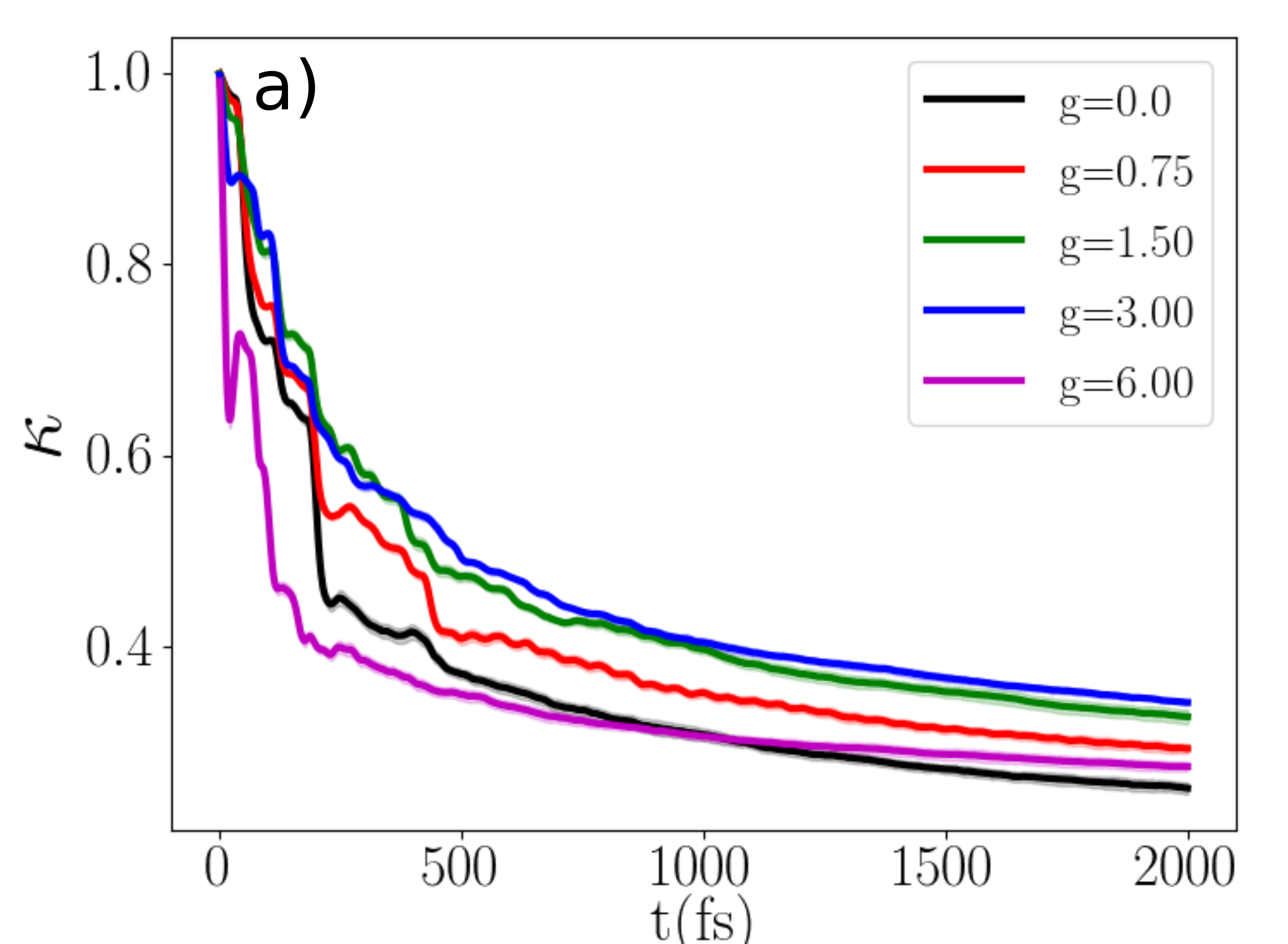}
\includegraphics[width=8cm]{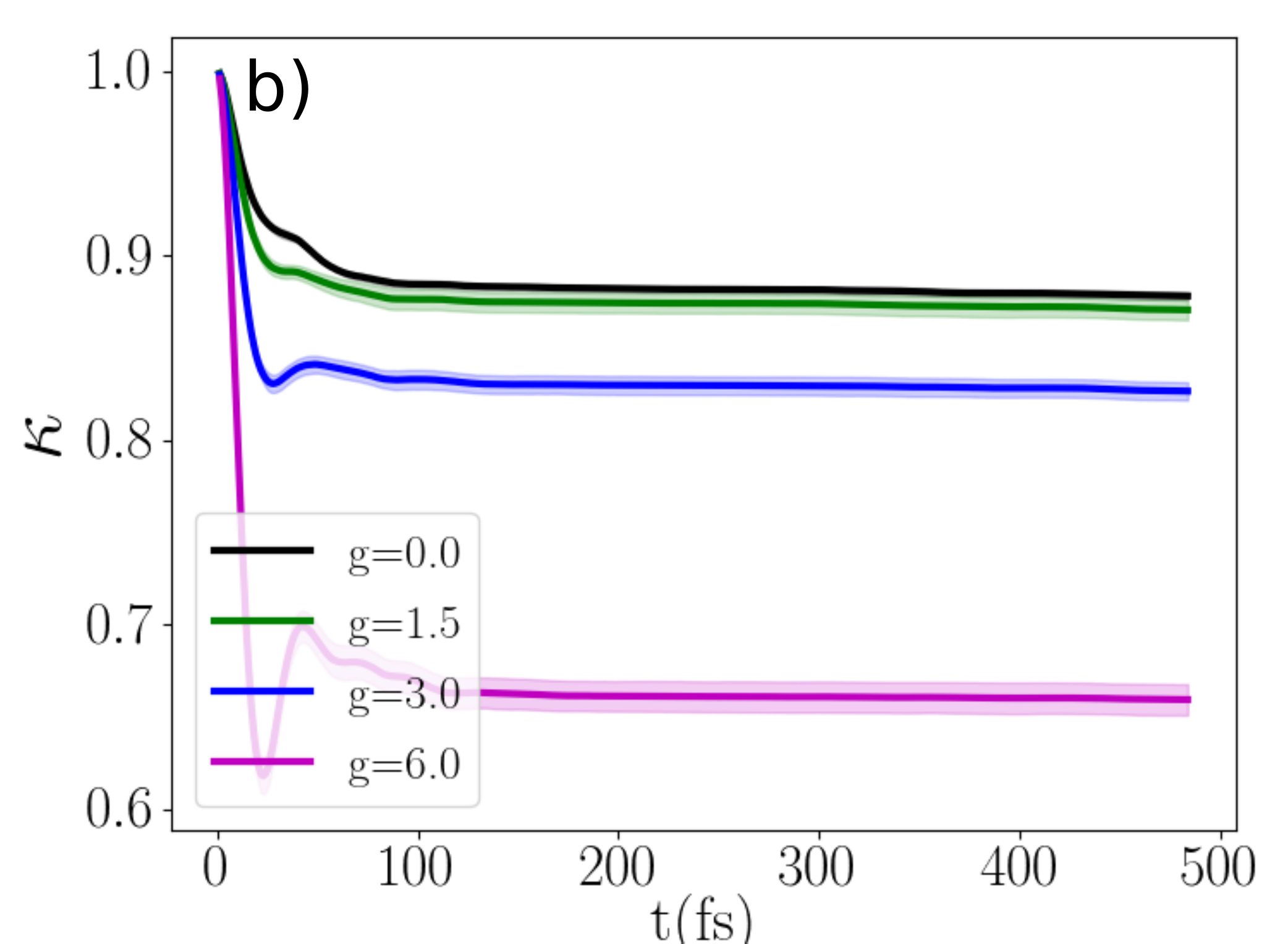}\\
\includegraphics[width=8cm]{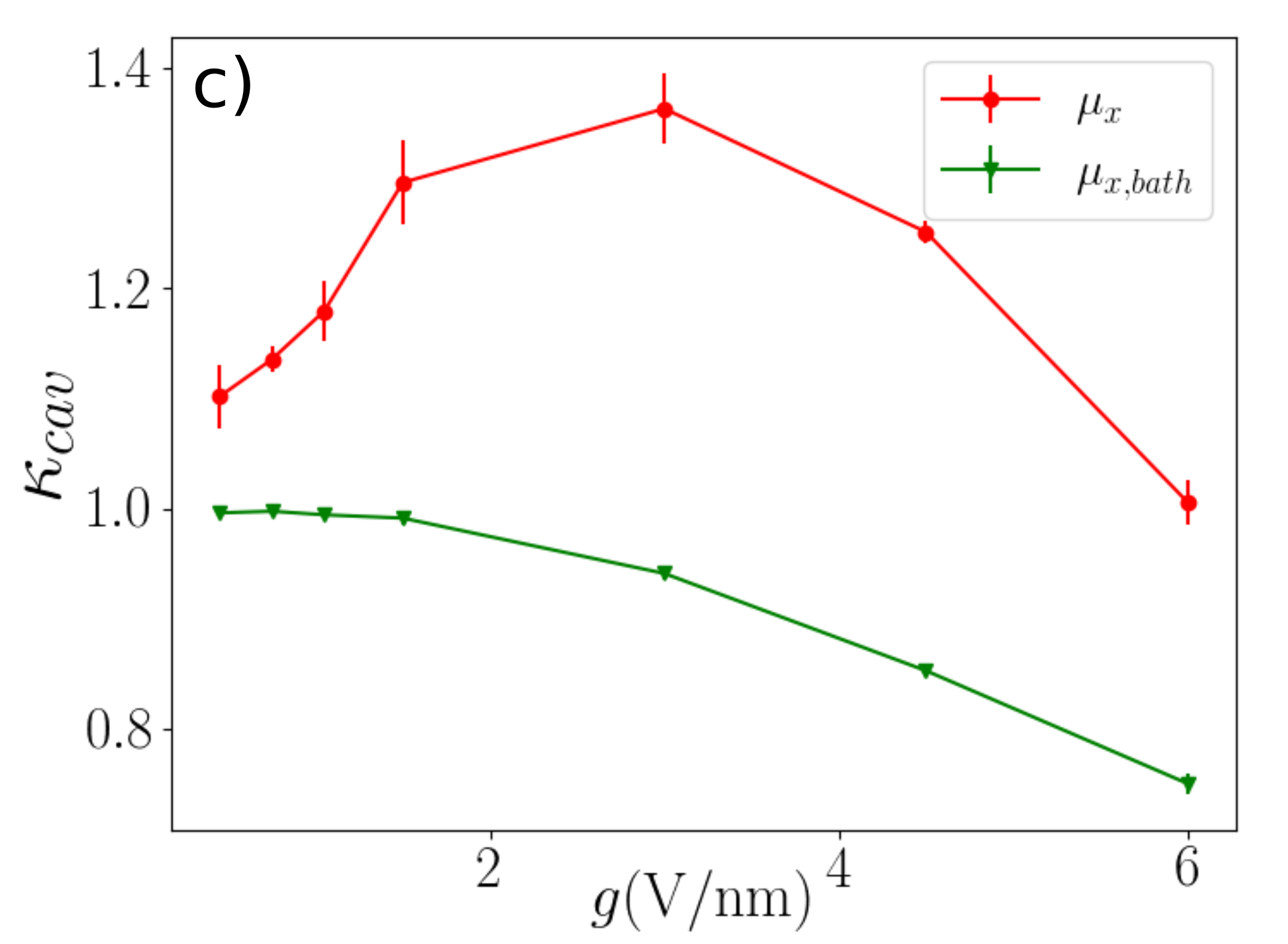}
\includegraphics[width=8cm]{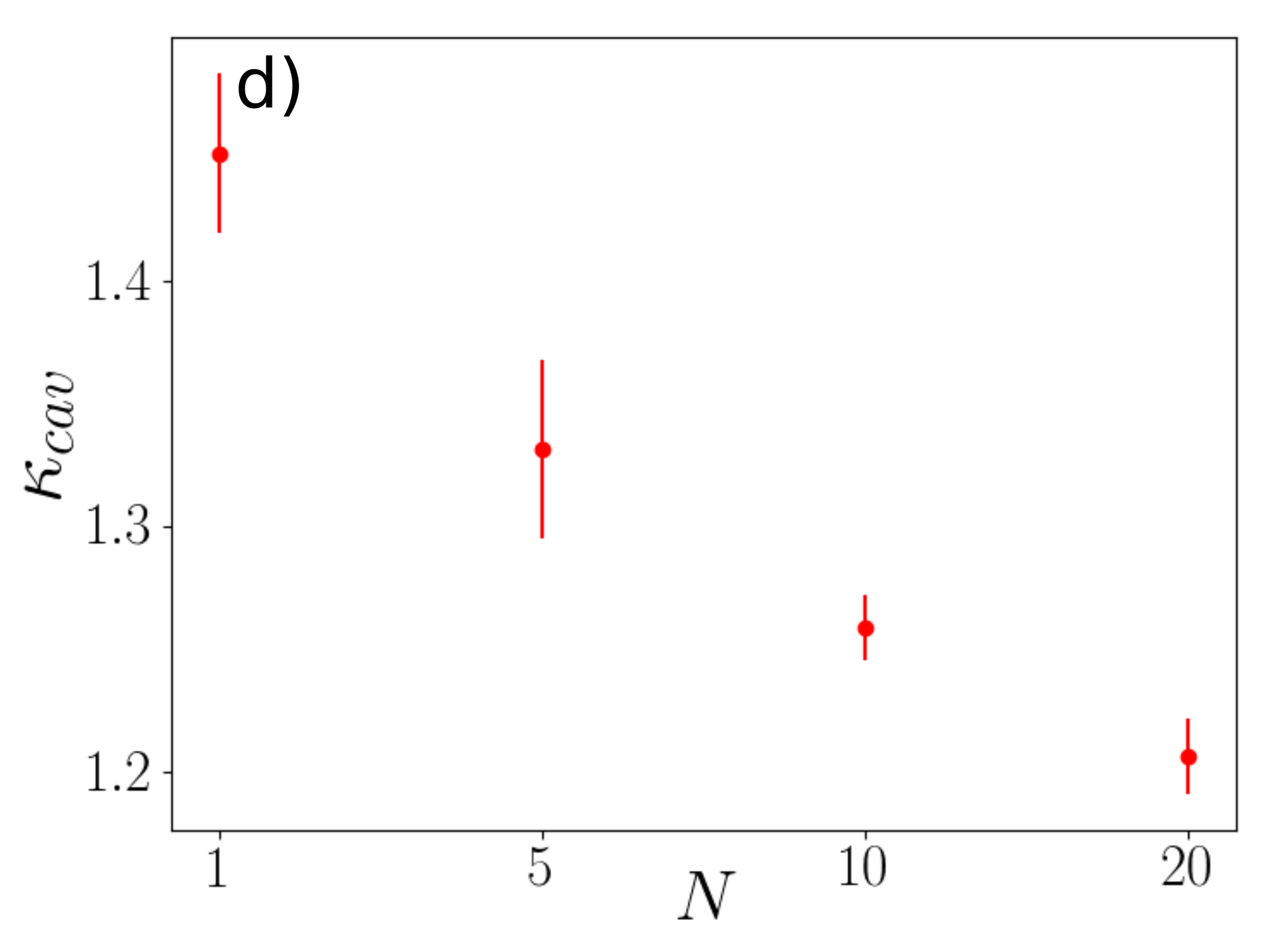}
\caption{\label{fig:k_rot}
a) Transmission coefficient $\kappa(t)$ for
various cavity-coupling strengths $g$ (V/nm) for $N=1$ and the cavity polarization
aligned with HONO's $x$-axis.
b) same as a) but with the HONO molecules coupled
to a bath (see SI). The shaded area on top of the solid lines indicates the standard deviation
of the average over trajectory runs.
c) Asymptotic $\kappa_{cav}=\kappa/\kappa_0$  for the curves in a) (red) and b) (green).
d) $\kappa_{cav}$ for increasing number of molecules at constant total
polaritonic coupling (see text for details).
}
\end{figure}

Let us consider a single HONO molecule coupled to a cavity mode with $x$ polarization
with respect to the molecular frame. In this case, the variation of the permenent dipole
is largest at the transition state (TS) of the reaction coordinate, $\tau^\ddagger\approx \pi/2$. 
Outside the cavity, $\kappa_0\approx 0.35$ at 300~K, the plateau value of the
black curve in Fig.~\ref{fig:k_rot}a. 
This relatively low transmission is caused by a slow rate of intra-molecular
vibrational energy redistribution (IVR) of the activated trajectories in the
underdamped regime.
As the cavity-coupling increases, one sees how the plateau value stabilizes at a
larger total transmission $\kappa$ in the red, blue and green curves.
The cavity accelerates the chemical reaction by increasing the total
transmission coefficient compared to $\kappa_0$, i.e.  $\kappa_{cav}>1$. This is
illustrated by the red trace in Fig.~\ref{fig:k_rot}c in the coupling regime
where $\kappa_{cav}$ increases,
and it is well understood within our theoretical framework: the cavity provides an
extra energy redistribution pathway for a system with a low-friction reaction
coordinate. Recrossing events are increasingly suppressed and the transmission
increases.
Nonetheless, as the cavity coupling to the torsion coordinate further increases,
a turning point is reached for $g>3$~V/nm. The amount of recrossing at the
barrier keeps increasing as well, thus finally reverting the trend and decreasing
the transmission again. This is the well-known Kramers turning point~\cite{hyn_86_149},
which, e.g., was predicted long ago for the isomerization of cyclohexane as a function of solvent
viscosity~\cite{kuh_88_3261}.
Figure~\ref{fig:k_rot}a illustrates its origin in the quick drop of $\kappa(t)$ at
short times for the strongest cavity coupling.

When adding an external bath to HONO (see SI for details), the regime of
validity of the GH theory is restored.
Figure~\ref{fig:k_rot}b shows how now $\kappa(t)$ quickly reaches the plateau value
within a few tens of femtoseconds, meaning that activated trajectories visit the region of the TS
only once or twice.
Since the plateau is reached quickly, the cavity
can only have a short-time effect close to the top of the barrier, where
it can increase the amount of recrossing thus reduce the transmission coefficient. 
As illustrated in Fig.\ref{fig:k_rot}c by the green trace, now $\kappa_{cav}<1$
and the chemical rate is reduced for all coupling strengths.
This is the regime captured in Refs.~\citenum{li_21_1315,yan_21_9531,man_22_014101}. 

\begin{figure}[t]
\includegraphics[width=8cm]{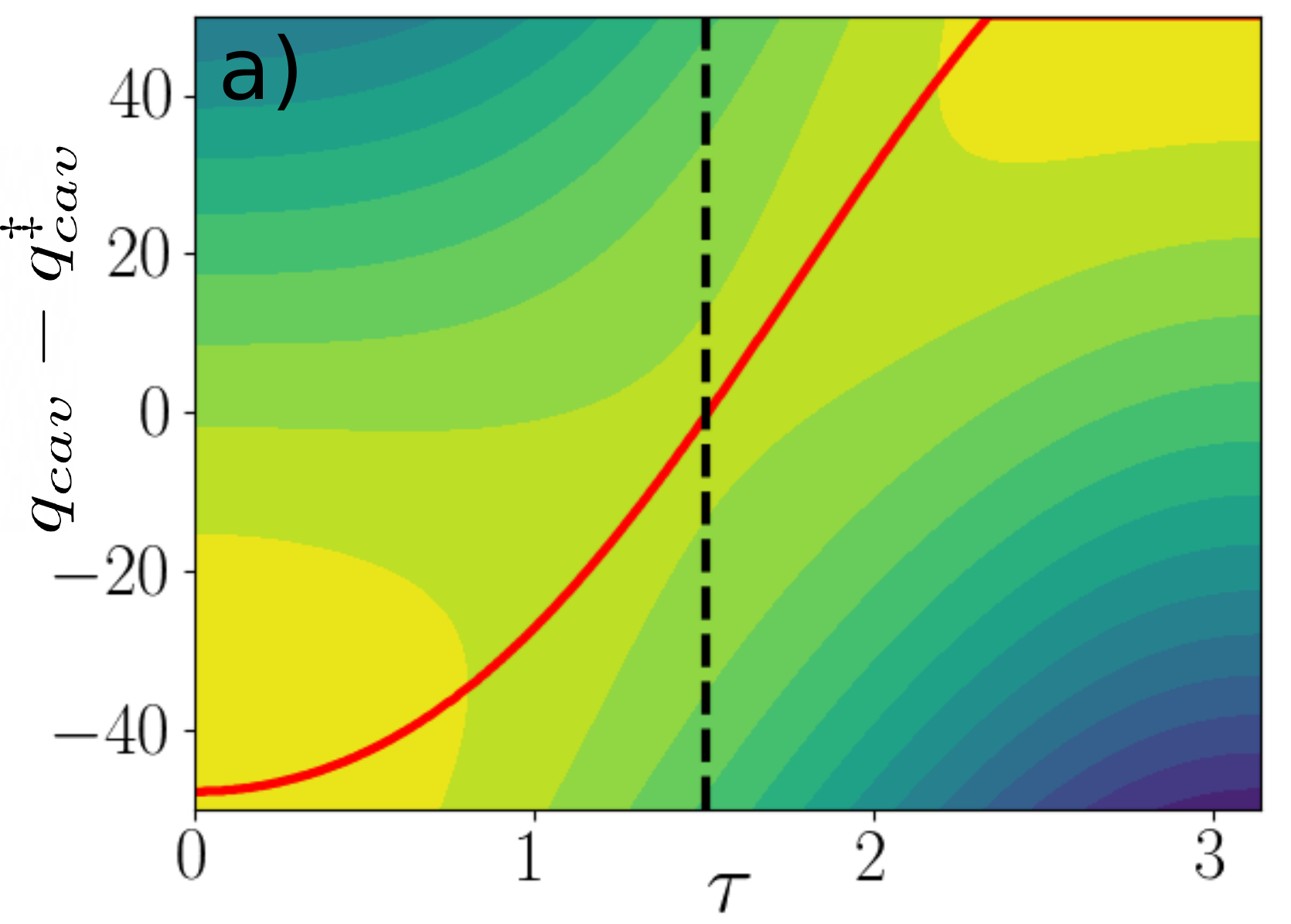}
\includegraphics[width=8cm]{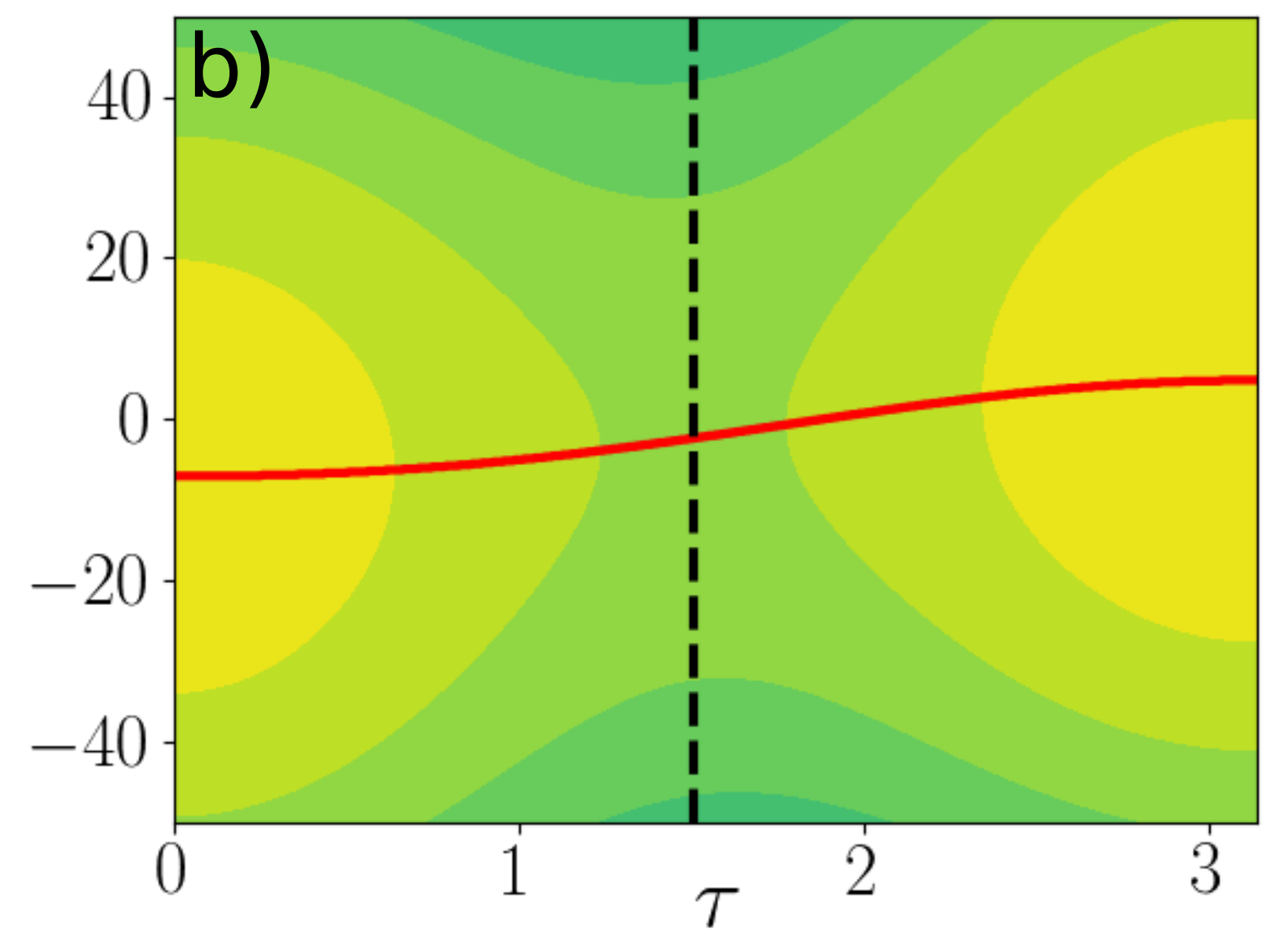}
\caption{\label{fig:PES_1} Potential energy surface cut for
  a) $1$ and b) $100$ HONO molecules
  as a function of the reaction coordinate $\tau$ and the cavity displacement
  $q_{cav}-q_{cav}^{\ddag}$. For $N=1$ the light-matter coupling is $g=8$~V/nm.
  The cavity coupling in b) is scaled by ${1/\sqrt{N}}$ to keep a constant overall
  light-matter interaction. The color levels start at $0$ for the lightest tone
  and increase in steps of $0.2$~eV. The red line indicates the
  minimum energy path. The vertical dashed line separates the \emph{cis} and \emph{trans} configurations.}
\end{figure}
A question mark remains still in connection with the collective VSC regime,
where most experiments reporting modifications of chemical rates in Fabry-Perot
configurations operate. 
To shed some light into this issue, we have performed trajectory
calculations of the transmission coefficient for an increasing number
of molecules $N$ coupled to the cavity, again without an extra bath.
The coupling per molecule is scaled as usual by a factor ${N}^{-1/2}$
as a means to keep the overall light-matter coupling constant~\cite{ven_18_253001}.
Starting from $N=1$, $g=1$~V/nm, and $\omega_{cav}=852$~cm$-1$,
one sees in Fig.~\ref{fig:k_rot}d
how, for increasing $N$, the cavity effect gradually fades away.
Responsible for the gradual trend $\kappa_{cav}\to 1$ is the decoupling of the reaction coordinate from
the cavity displacement with increasing $N$, as seen by comparing
the curvature of the minimum energy path (MEP) in
Figs.~\ref{fig:PES_1}a, $N=1$, and \ref{fig:PES_1}b, with $N=100$.
This reduction of the MEP curvature as $N$ increases,
and thus the reduced friction caused by the cavity, implies that
in the large $N$ limit the cavity is not able to ``cage'' the TS and induce a
decrease of the transmission coefficient through this mechanism.

\begin{figure}
  \includegraphics[width=8cm]{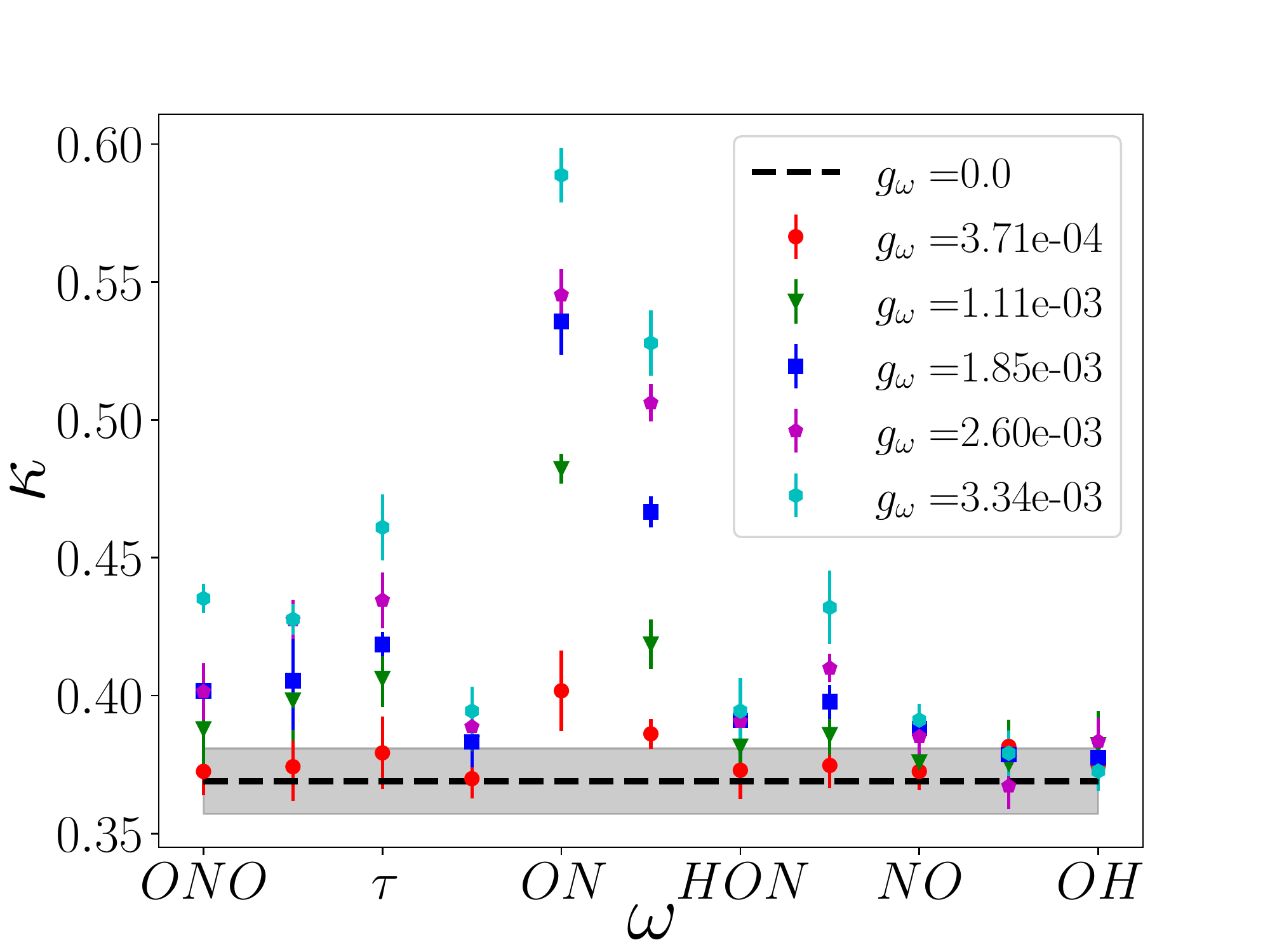}
  \caption{\label{fig:k_tot} Transmission coefficient $\kappa$
      for various coupling strengths $g_\omega=g\mu_x^{\ddag}/\omega_{cav}$
  as a function of $\omega_{cav}$. Vertical bars represent standard
  deviations over the run trajectories. $\omega_{cav}$ is chosen to
  be resonant with fundamental modes of molecule: $\omega_{ONO}=609 (cm^{-1})$,
  $\omega_{\tau}=640 (cm^{-1})$, $\omega_{ON}=852 (cm^{-1})$, $\omega_{HON}=1263
  (cm^{-1})$, $\omega_{NO}=1641 (cm^{-1})$, $\omega_{OH}=3426 (cm^{-1})$ and
  with the average of every consecutive pair. The black dashed line indicates
  $\kappa_0$ outside the cavity with its standard deviation.
  }
\end{figure}
Finally, we address the question of sharp resonant effects, meaning when the
modification of chemical rates is particularly pronounced at specific cavity
frequencies.
Through trajectory calculations it has been observed that the
outcome of reactive events can depend on the resonance between the cavity
and vibrational modes of the molecule, but a link to the actual modification of
chemical rates has not been established~\cite{sid_20_7525,sch_21_}.
Our simulations of the transmission coefficient in the
underdamped, slow IVR regime reveal
sharp resonances in the rate constant effect as a function of $\omega_{cav}$.
As already discussed, in this regime the effect of the cavity is to
introduce extra energy redistribution pathways, whereby the effect at short
times while passing the TS barrier region is not so important. Thus, when the cavity is
resonant with a vibrational mode that happens to be strongly coupled to the
reaction coordinate, the enhancement of the rate is more prominent. As seen in
Fig.~\ref{fig:k_tot}, $\kappa_{cav}\approx 2$ when $\omega_{cav}$ is resonant
with the O-N stretching mode at 852~cm$^{-1}$.
This is not surprising. It is well-known that the O-N stretch is strongly
coupled to the torsion coordinate in HONO~\cite{ric_04_1306, ric_04_6072}:
Selective laser excitations of this mode result in an enhanced probability of
isomerization out of equilibrium~\cite{ric_07_164315}.
This brings us to the observed cavity catalysis of a unimolecular dissociation
in solution by the Ebbesen group~\cite{lat_19_10635}.  A plausible explanation
is that the strong resonance of the cavity mode with a carbonyl group could
stabilize the hot nascent products inside the solvent pocket, in this way
preventing recrossing events at early times after passage over the TS.
Further studies will be required to test our hypothesis.

Let us recapitulate and place our findings in the context of the current
literature on thermal rate models in cavities.
The regime considered in model studies is the one in which the cavity is strongly coupled
to the reaction coordinate, while reactants and products are
strongly damped by a bath~\cite{li_21_1315}
or effectively by a short propagation time that hinders recrossing~\cite{gal_19_021057}.
In this case, the reaction pathway becomes curved, as seen in
the Shin-Metieu and other 1D models~\cite{gal_19_021057,li_21_1315,man_22_014101}.
The regime in which the coupling is strong at the barrier top and the reactants and
products are damped by a bath is well captured
by the Grote-Hynes (GH) theory~\cite{hyn_86_149} and has been the basis of theoretical proposals
for the reduction of chemical rates in cavities~\cite{li_21_1315,man_22_014101,yan_21_9531}.
In this regime there are no sharp resonance effects,
although there is a continuous dependence of $\kappa$ on $\omega_{cav}$ related
to the frequency of the inverted reaction barrier~\cite{man_22_014101}.
Similarly to our findings, existing theories report a reduction of the
transmission coefficient in the collective regime as $N$ increases (cf. Fig. 2a
in Ref.~\citenum{man_22_014101}).
Reference~\citenum{yan_21_9531} attributes the persistence of the modification of
chemical rates in the large $N$ collective regime to the assumption that ``the polariton
state is thermally activated to yield collective barrier
crossing''~\cite{yan_21_9531}. While a coherent superposition of activated
complexes may display this behavior, a physical mechanism by which a
polaritonic system spontaneously results in coherence in the dark and
under thermal equilibrium of the material and
photonic modes, has not been put forward.

Concluding, our main contribution has been to identify dynamical effects played by the
cavity in the low-friction (or underdamped) regime of the reaction coordinate.
In this regime, the cavity effect is twofold:
(1) It \emph{accelerates} the chemical rate by increasing the friction compared to
the cavity-free system. This reduces the recrossing due to trajectories that,
otherwise, would visit reactants and products several times, thus increasing the
transmission coefficient compared to $\kappa_0$.  As the cavity coupling keeps
increasing, the overall increased friction can introduce again more coupling at
the barrier and the trend can overturn. This is the well-known Kramers turnover
situation, and this regime can exist in the
condensed phase~\cite{hyn_86_149,kuh_88_3261}.
(2) In the low-friction regime, sharp resonant effects are possible.  These are
related to the new IVR pathways offered by the cavity. In the products-side,
they dissipate energy from the nascent hot products. In the reactants-side, the
resonances funnel energy into the nascent activated complexes. Numerically, the
former is captured by trajectories starting towards the products-side and being
effectively captured there.  The latter is captured by the trajectories
initially moving towards reactants and being effectively captured as well. If this reactants-side
capture would be ineffective, they would be counted as products but with a
negative contribution to the flux, in this way lowering the transmission
(cf. Eq.~\ref{eq:kt}).
Finally, when a bath is added to the HONO molecule and the overall friction is
sufficiently increased, the model reverts to the already known GH regime where the cavity
only affects the recrossings at the top of the reaction barrier.
Our findings shed important new light onto the question of cavity-modified reactivity. However,
it still remains for future work to better understand how these cavity effects
can survive in actual liquid phases and in the collective regime for
truly macroscopic numbers of molecules. It is plausible that the
more detailed answers lie beyond models of
independent molecules and may require studies of the transmission
coefficient with full consideration of environmental effects in the
bulk~\cite{li_21_15533}.


\ifx\mcitethebibliography\mciteundefinedmacro
\PackageError{achemso.bst}{mciteplus.sty has not been loaded}
{This bibstyle requires the use of the mciteplus package.}\fi

\end{document}


\section{Calculation of the transmission coefficient}

    Starting from the general definition of the forward reaction rate
    \begin{align}
    \label{eq:kt}
    K(t) &= x_{cis}^{-1}
    \langle \dot{\tau}(0)\,
            \delta[\tau(0)-\tau^{\ddag}]\,
            \theta[\tau(t)] \rangle
    \end{align}
    and its TST limit
    \begin{align}
        \label{eq:ktst}
        K_{TST} &= x_{cis}^{-1}
        \langle \dot{\tau}(0^+)\,
                \delta[\tau(0)-\tau^{\ddagger}] \rangle,
        \end{align}
    which integrates only over the forward flux assuming it all
    leads to products, the transmission coefficient
    \begin{align}
        \label{eq:kappa}
        \kappa(t) &= \frac{\langle \dot{\tau}(0)\,\delta[\tau(0)-\tau^{\ddagger}]\,\theta[\tau(t)] \rangle}
                          {\langle \dot{\tau}(0^+)\,\delta[\tau(0)-\tau^{\ddagger}] \rangle}
    \end{align}
    can be defined.
    $\kappa$ contains all dynamical information that escapes the zero-recrossing, $\kappa=1$
    TST limit. The rate $K = \kappa K_{TST}$, where $\kappa\equiv \lim_{t\to\infty}\kappa(t)$,
    is the exact classical forward rate (if it exists: for very low barriers between two
    meta-stable configurations $\kappa$ tends quickly to 0 and no rate can be defined).

    $\kappa(t)$ is calculated as follows:
    For every set of parameters ($g$, $\omega_{cav}$, $N$) considered, a
    Metropolis sampling in conguration space samples the canonical distribution
    with probability proportional to $\exp{-\beta V(\vec{R}^\ddagger)}$, with
    $\beta^{-1}=k_B T$ and $T=300$~K. In $\vec{R}^\ddagger$, the torsion
    coordinate is constrained to its transition state (TS) value $\tau^\dagger$,
    and the corresponding configurational space is therefore 5-dimensional.
    In the Metropolis sampling, the step size of the random moves is adjusted in
    an initial burn-in phase to attain an acceptance ratio
    of 40\%.
    To reduce the correlation of contiguous samples in the generated Markov
    chain, only one in every 5 samples is kept.
    Finally, $5\times 10^5$ initial configurations are collected
    and split in 5 independent
    batches.
    The corresponding velocities are initialized from a Maxwell-Boltzmann
    distribution.
    The classical equations of motion of the system plus cavity are integrated
    in the Newtonian formulation using the velocity-Verlet algorithm with a
    time-step of 1~fs.
    The standard deviation of each run is determined over the averages of the
    5 independent batches.

\section{Transition state theory rates}

    Although TST rates can be directly evaluated using Eq.~\ref{eq:ktst}, we
    find it more convenient to use Eyring's equation
    \begin{align}
          \label{eq:ktsteyr}
          k_{TST} = \frac{1}{\beta h}\frac{Z^{\ddag}}{Z^{R}}e^{-\beta \Delta E}
    \end{align}
    where  $Z^{\ddag}$ and $Z^{R}$ correspond to the partition function of
    transition state (TS) and reactant, respectively.
    The partition functions are
    evaluated as usual by optimizing the \emph{cis} and TS
    stationary point geometries for each set of parameters and constructing and
    diagonalizing the Hessian matrix to obtain separable normal modes.
    Finally, the quantum mechanical partition functions for the
    oscillators in the reactants and TS are evaluated
    using the well-known textbook expressions. $\Delta E$ in Eq.~\ref{eq:ktsteyr}
    corresponds to the difference of zero-point energies between reactants and TS.

    \begin{figure}[h]
        \includegraphics[width=8cm]{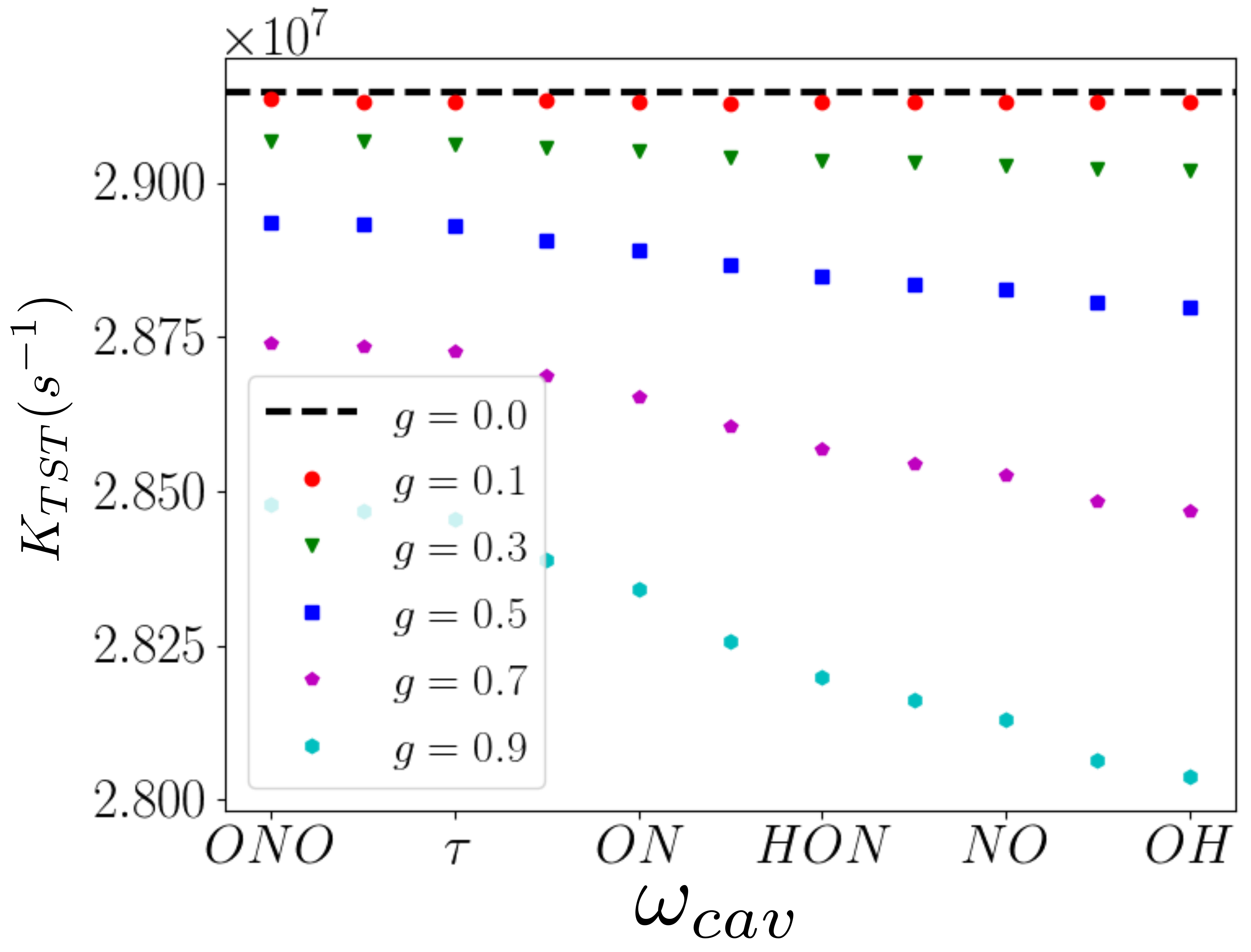}
        \caption{Transition state theory rates calculated with Eyring's
        equation. The frequencies for $\omega_{cav}$ are taken to be resonant
        with the harmonic vibrational frequencies of HONO. Their values
        are given in the caption to Fig. 4 in the main text.}
    \end{figure}
    
\section*{HONO permanent dipole}

    The permanent dipole moment of HONO has been evaluated at the MP2
    level of theory using the atomic basis 6-311g++(d,p).

    \begin{figure}[h]
    	\includegraphics[width=10cm]{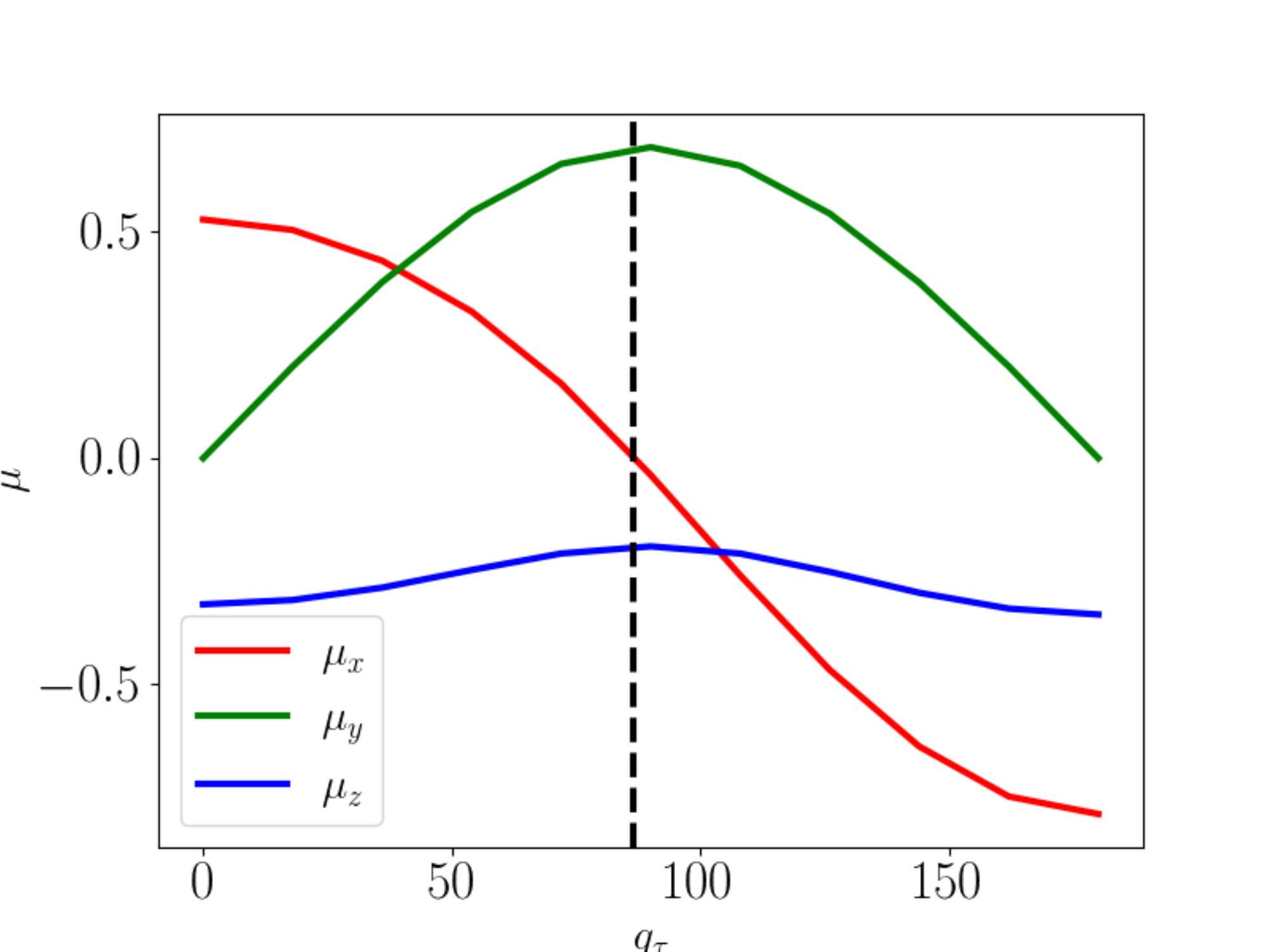}
    	\caption{\label{fig:mu_cis} Permanent dipole of the HONO molecule in atomic units
        as a function of the torsion coordinate $\tau$. The axes are referred to
        the molecular frame axis in Fig. 1 of the main text.}
    \end{figure}

\section*{Calculations with a bath}
	
	For the calculations with a bath coupled to the reaction coordinate,
	the bath Hamiltonian is modeled as
	\begin{align}
		\hat{H}_{bath} = \sum_{k}\frac{p_{k}^2}{2m_{k}} + \frac{m_{k}\omega_{k}^2}{2}\left[q_k+\frac{c_k}{m_{k}\omega_{k}^2}(R_{rxn}-R^{\ddagger}_{rxn}) \right]^2,
	\end{align}
	which describes the interaction between the reaction coordinate ($R_{rxn}$) of the system and bath modes,
	which linearly grows with displacements from the TS.
	$q_k$ denotes the coordinate of the $k$-th bath mode with a conjugate momentum $p_k$ and mass $m_k = 1836$ a.u..
	The coupling constants $c_k$ and the frequencies $\omega_k$ are chosen to represent
	an Ohmic spectral density
	\begin{align}
		J(\omega) = \frac{\pi}{2}\sum_{k}\frac{c_k^2}{m_k\omega_k}\delta(\omega - \omega_k) = \frac{\pi}{2}\hbar\xi\omega e^{-\omega/\omega_p},
	\end{align}
	where $\omega_p$ is set to be $170.6$ meV and $\xi$ is a unit-less factor ($\xi = 1.74$ in our calculation).
	Hence, one can derive $\omega_k$ and $c_k$ from
	\begin{align}
		\omega_k = -\omega_p ln(1-k(\frac{\omega_N}{\omega_p})),
	\end{align}
	and
	\begin{align}
		c_k=\sqrt{\xi\hbar\omega_N m_k}\omega_k,
	\end{align}
	where $\omega_N = \frac{\omega_p}{N}(1-e^{-\omega_m/\omega_p})$.
	For production calculations we used $N=20$ bath modes, but we tried with up to $N=80$ modes to
	check for convergence and the system's observables remained unchanged.
	$\omega_N$ denotes the largest frequency in the bath ($\omega_N = 5\omega_p$).